\documentclass{svmult}

\usepackage{graphicx}    

\def\beq{\begin{equation}}
\def\eeq{\end{equation}}

\def\bea{\begin{eqnarray}}
\def\eea{\end{eqnarray}}
\def\ba{\begin{array}}                  
\def\ea{\end{array}}

\def\ve{\varepsilon}

\begin{document}

\title*{Towards 
a NNLO  calculation
in hadronic heavy hadron
 production }
\author{J\"urgen G.\ K\"orner \inst{1, a},  Zakaria\ Merebashvili
 \inst{2, b} \and Mikhail Rogal
 \inst{1, c, *} }
\institute{Institut f\"ur Physik,  Johannes Gutenberg-Universit\"at, D55099 - Mainz, Germany
\and High Energy Physics Institute,
Tbilisi State University, 380086 Tbilisi, Georgia  
}
\maketitle

\section { Introduction}
\label{sec:1}

The full next-to-leading order (NLO) corrections to the hadroproduction of heavy 
flavors have been completed in 1988 \cite{Dawson:1988,Been}. They 
have raised the leading order (LO) estimates \cite{LO:1978} but were 
still below  the experimental results (see e.g. \cite{Italians}).
In a recent analysis theory moved closer to experiment \cite{Italians}. 
A  large uncertainty in the NLO calculation results from the freedom in the 
choice of the renormalization and factorization scales.
The dependence on the factorization and renormalization scales  is expected to be greatly reduced at next-to-next-to-leading order(NNLO). This
reduces the theoretical uncertainty. Furthermore, one may hope that  there is yet  better agreement between 
theory and experiment at NNLO.
\renewcommand{\thefootnote}{a}
\footnotetext{koerner@thep.physik.uni-mainz.de}
\renewcommand{\thefootnote}{b}
\footnotetext{zaza@thep.physik.uni-mainz.de}
\renewcommand{\thefootnote}{c}
\footnotetext{rogal@thep.physik.uni-mainz.de}
\renewcommand{\thefootnote}{*}
\footnotetext{Presented by M. Rogal at the  9th Adriatic Meeting - Central European Symposia `` Particle physics and the universe '', Dubrovnik, Croatia, 2003; to be published in the Proceedings. }
\begin{figure}
\center
\includegraphics[height=4cm]{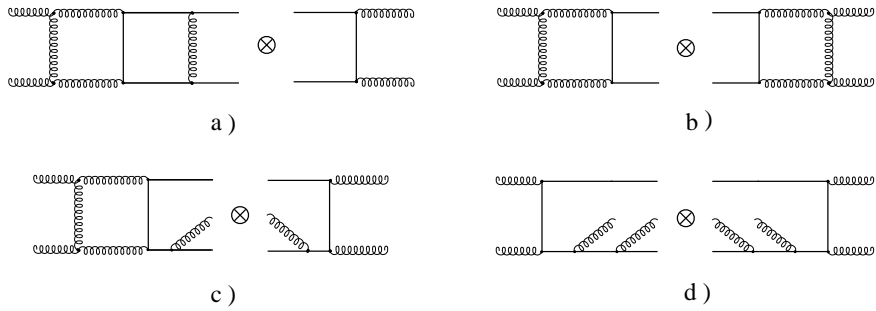}
\caption{Exemplary diagrams  for the  NNLO calculation of heavy hadron production}
\label{nnlo}       
\end{figure}

In Fig.~\ref{nnlo} we show one generic diagram each for the four classes
of contributions that need to be calculated for the NNLO corrections
to hadroproduction of heavy flavors. They involve the two-loop
contribution (Fig.~\ref{nnlo}a), the loop-by-loop contribution (Fig.~\ref{nnlo}b), the one-loop
gluon emission contribution (Fig.~\ref{nnlo}c) and, finally, the two gluon
emission contribution (Fig.~\ref{nnlo}d). An interesting subclass of the diagrams 
in Fig.~\ref{nnlo}c are those  diagrams where  the outgoing gluon is attached directly to the
 loop. One then has a five-point function which has to be calculated up to ${\cal O}(\ve^{2})$.

 In our work we have concentrated on the
loop-by-loop contributions exemplified by Fig.~\ref{nnlo}b. Specifically, working in the framework 
of  dimensional regularization, we have calculated  ${\cal O}(\ve^2)$ results for
all scalar one-loop one-, two- and  three-point integrals that are
needed in the calculation of hadronic heavy flavour production. Work on the relevant four-point integrals is in progress.  
The integrations were generally done by writing down the Feynman parameter representation 
for the corresponding integrals, integrating over Feynman parameters up to 
the last remaining integral, expanding the integrand of the last remaining 
parametric integral in terms of $\ve$ and doing the last parametric integration on 
the coefficients of the expansion.
Because the one-loop integrals exhibit infrared (IR)/collinear (M)
singularities up to ${\cal O}(\ve^{-2})$ one needs to know the one-loop 
integrals up to ${\cal O}(\ve^2)$ because the one-loop contributions appear 
in product form in the loop-by-loop contributions. It is clear that the spin 
algebra and the calculation of tensor integrals in the 
one-loop contributions also have to be done up to ${\cal O}(\ve^2)$. 
This work is in progress.

   Due to lack of space we can only present a few exemplary results. 
Since the four-point functions are the most difficult  we concentrate on them.
 
\section{Four-point functions}
\begin{figure}
\center
\includegraphics[height=2cm]{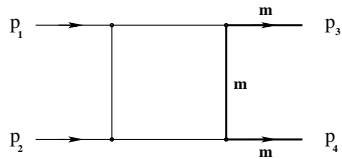}
\caption{One-loop box with one internal massive propagator }
\label{box}     
\end{figure}

As a sample calculation we  discuss the $(0,0,m,0)$-box with one massive propagator depicted in Fig.~\ref{box}.
As explained before one needs to calculate each one-loop integral up to ${\cal O}(\ve^{2})$ in order to obtain the finite  terms in the loop-by-loop contributions.
The box integral Fig.~\ref{box} is represented by the integral 
\newpage
\begin{eqnarray}
\lefteqn{D(-p_2,p_4,p_3,0,0,m,0)=} \nonumber   \\
        &&
             \mu^{2\ve} \int \frac{d^nq}{(2\pi)^n}
           \frac{1}{(q^2)(q-p_{2})^2[(q-p_{2}+p_{4})^2-m^2]
                    (q-p_{2}+p_{4}+p_{3})^2}\quad , \nonumber
\end{eqnarray}
where $p_{1}$, $p_{2}$, $p_{3}$ and $p_{4}$ are external  momenta with $p_{1}^{2}=p_{2}^{2}=0,\, p_{3}^{2}=p_{4}^{2}=m^{2}$ and $n=4-2\ve$ is the dimension of space-time.

The $\ve^{-2}$, $\ve^{-1}$ and $\ve^{0}$ coefficients have been known for some time \cite{Dawson:1988,Been} and will not be listed here.
We define Mandelstam-type variables by $s\equiv (p_1+p_2)^2,\, t\equiv (p_1-p_3)^2-m^2$ 
and $\beta=(1-4m^2/s)^{1/2},\, x=(1-\beta)/(1+\beta)$.

For the real part of the ${\cal O} (\ve)$ term we obtain:\\
\\
$
\frac{iC_{\ve}(m^{2})}{12s t} \ve {\Big [ } 6\, \ln^{3}\frac{s}{m^2} + 20\, \ln^{3}\frac{-t}{m^2} + \ln^{3} x + 
  6\, \ln^{2} x\, \ln\frac{-\left(s + 2\, t - s\, \beta \right)}{2\, m^2} \\- 
  12\, \ln x\, \ln^{2}\frac{-\left(s + 2\, t - s\, \beta \right)}{2\, 
              m^2}) + 
  8\, \ln^{3}\frac{-\left(s + 2\, t - s\, \beta \right)}{2\, m^2} \\- 
  12\, \ln^{2}\frac{-t}{m^2} \, \left(2\, \ln(-1 - \frac{t}{m^2}) + 
        3\, \ln x -
        \ln\frac{-\left(s + 2\, t - s\, \beta \right)}{2\, m^2} + 
        4\, \ln\frac{s + 2\, t + s\, \beta}{2\, m^2} \right)\\ - 
  3\, \ln^{2}\frac{s}{m^2}\, \left(6\, \ln\frac{-t}{m^2} + 3\, \ln x + 
        2\, \ln\frac{-\left(s + 2\, t - s\, \beta \right)}{2\, m^2} + 
        4\, \ln\frac{s + 2\, t + s\, \beta}{2\, m^2} \right) + 
  12\, \ln \frac{s}{m^2} + 24\, \ln \frac{-t}{m^2} + 
  24\, \ln \frac{s}{m^2}\, \ln \frac{-t}{m^2} + 48\, 
Li_3(\frac{m^2 + t}{t}) + 48\, Li_3(1 + \frac{t}{m^2}) + 24\, Li_3(-x) - 24\, 
Li_3(\frac{-2\, 
          m^2 + t\, \left(-1 + \beta \right)}{t\, \left(1 + \beta \right)}) - 
  48\, Li_3(\frac{2\, m^2 + t - t\, \beta}{2\, m^2}) + 24\, 
Li_3(\frac{2\, m^2 + t - t\, \beta}{m^2\, \left(1 + \beta \right)})\\ - 24\, 
Li_3(\frac{2\, m^2 + t - t\, \beta}{t - t\, \beta}) - 48\, 
Li_3(\frac{2\, m^2}{2\, m^2 + t + t\, \beta}) + 24\, 
Li_3(- \frac{m^2\, \left(-1 + \beta \right)}{2\, 
              m^2 + t + t\, \beta} )\\ - 24\, 
Li_3(\frac{t\, \left(-1 + \beta \right)}{2\, m^2 + t + t\, \beta}) - 24\, 
Li_3(\frac{2\, m^2 + t + t\, \beta}{t + t\, \beta}) - 
  6\, \ln\frac{-t}{m^2} \,{ \Big(}4 + 
        2\, \ln^{2}(-1 - \frac{t}{m^2}) - \ln^{2} x - 
        4\, \ln x\, \ln\frac{-\left(s + 2\, t - s\, \beta \right)}{2\, 
                m^2} + 
        6\, \ln^{2}\frac{-\left(s + 2\, t - s\, \beta \right)}{2\, m^2} \\- 
        4\, \ln(-1 - \frac{t}{m^2})\, \left( \ln x + \ln\frac{-\left(s + 2\, 
                          t - s\, \beta \right)}{2\, m^2} +
              \ln\frac{s + 2\, t + s\, \beta}{2\, m^2} \right) + 8\, 
      Li_2(1 + \frac{t}{m^2}) \\+ 8\, 
      Li_2(\frac{t\, \left(-1 + \beta \right)}{2\, m^2}) + 4\, 
      Li_2(\frac{m^2\, \left(-1 + \beta \right)}{\left(m^2 + 
                        t \right)\, \left(1 + \beta \right)}) - 4\, 
      Li_2(\frac{\left(m^2 + 
                        t \right)\, \left(-1 + \beta \right)}{m^2\, \left(1 + \
\beta \right)}) \\ + 8\, 
      Li_2(\frac{2\, m^2}{2\, m^2 + t + t\, \beta}) - 12\, \zeta(2) {\Big)} - 
  12\, \ln x\, \zeta(2) + 
  48\, \ln\frac{-\left(s + 2\, t - s\, \beta \right)}{2\, m^2}\, \zeta(2) - 
  6\, \ln\frac{s}{m^2}\, {\Big(}2 + 4\, \ln^{2}\frac{-t}{m^2}  - \ln^{2} x - 
        2\, \ln x\, \ln\frac{s + 2\, t + s\, \beta}{2\, m^2} - 
        4\, \ln\frac{-t}{m^2} \, {\big(}-1 + \ln x + \ln\frac{-\left(s + 2\, 
                          t - s\, \beta \right)}{2\, m^2} + 
              2\, \ln\frac{s + 2\, t + s\, \beta}{2\, m^2} {\big)} - 4\, 
      Li_2(\frac{t\, \left(-1 + \beta \right)}{2\, m^2}) - 4\, 
      Li_2(\frac{2\, m^2}{2\, m^2 + t + t\, \beta}) + 2\, \zeta(2) {\Big)} - 
  36\, \zeta(3)  {\Big ]}
$,\\ where $C_{\ve}(m^2)\equiv\frac{\Gamma(1+\ve)}{(4\pi)^2} \left(\frac{4\pi\mu^2}{m^2}\right)^\ve$.\\

One also needs the imaginary part of the $(0,0,m,0)$-box since the total contributions from the loop-by-loop contribution contains also imaginary parts  via $|A|^{2}=(Re A)^{2}+(Im A)^{2}$. Note , however, that the imaginary part is only needed up to ${\cal O}(\ve)$  since the IR/M singularities in the imaginary parts of the one-loop contributions are of  ${\cal O}(\frac{1}{\ve})$ only. For the ${\cal O}(\ve)$ absorptive (imaginary) part we obtain:\\
\\
$
- \frac{iC_{\ve}(m^{2})}{4s t}\ve \pi {\Big[}3\, \ln^{2} \frac{s}{m^2} - 
  4\, \ln \frac{s}{m^2}\, \ln \frac{-t}{m^2}  - 
  8\, \ln^{2}\frac{-t}{m^2}  - 
  2\, \ln \frac{s}{m^2}\, \ln x + 
  4\, \ln \frac{-t}{m^2}\, \ln x+ 3\, \ln^{2} x - 
  4\, \ln\frac{s}{m^2}\, \ln \frac{-\left(s + 2\, 
              t - s\, \beta \right)}{2\, m^2} - 
  4\, \ln x\, \ln \frac{-\left(s + 2\, t - s\, \beta \right)}{2\, m^2} + 
  4\, \ln^{2} \frac{-\left(s + 2\, t - s\, \beta \right)}{2\, m^2} - 
  4\, \ln \frac{s}{m^2}\, \ln \frac{s + 2\, t + s\, \beta}{2\, m^2} + 
  8\, \ln \frac{-t}{m^2} \, \ln \frac{s + 2\, 
          t + s\, \beta}{2\, m^2} + 
  4\, \ln x\, \ln \frac{s + 2\, t + s\, \beta}{2\, m^2} + 8\, 
Li_2(\frac{t\, \left(-1 + \beta \right)}{2\, m^2}) + 8\, 
Li_2(\frac{2\, m^2}{2\, m^2 + t + t\, \beta}) + 8\, \zeta(2)
{\Big ] }.
$\\

The  $\ve^{2}$--results for the $(0,0,m,0)$-box  are too lengthy to fit into this report. They will be presented in a forthcoming publication \cite{KMR}. A new feature of the $\ve^{2}$-- contributions is that the result can no longer be expressed in terms of logarithms and polylogarithms . They involve more general functions - the multiple polylogarithms introduced by Goncharov in 1998 \cite{Gonch}. A multiple polylogarithm is represented by
\begin{eqnarray}
Li_{m_{k},...,m_{1}}(x_{k},...,x_{1})=\int \limits_{0}^{x_{1}x_{2}...x_{k}} \left( \frac{dt}{t} \circ \right)^{m_{1}-1} 
\frac{dt}{x_{2}x_{3}...x_{k}-t} \circ 
\nonumber \\ \left( \frac{dt}{t} \circ \right)^{m_{2}-1}  \frac{dt}{x_{3}...x_{k}-t} \circ ...  \circ  \left( \frac{dt}{t} \circ \right)^{m_{k}-1}   \frac{dt}{1-t}\, ,\nonumber
\end{eqnarray}
where the iterated integrals are defined by
\begin{eqnarray}
\int \limits_{0}^{\lambda} \frac{dt}{a_{n}-t}\circ ...\circ  \frac{dt}{a_{1}-t}=\int \limits_{0}^{\lambda} \frac{dt_{n}}{a_{n}-t_{n}}
\int \limits_{0}^{t_{n}} \frac{dt_{n}}{a_{n-1}-t_{n-1}} \times...\times \int \limits_{0}^{t_{2}}\frac{dt}{a_{1}-t_{1}}.\nonumber
\end{eqnarray}
Besides the scalar $(0,0,m,0)$-box one also needs to calculate the scalar $(0,0,m,m)$ and $(m,m,0,m)$-boxes. Work is in progress on the calculation of these boxes.

\section{Summary}
We  have  reported on the results of an ongoing calculation of the $\ve$-expansion of the scalar one-loop integrals up to ${\cal O}(\ve^{2})$ that are needed for the NNLO calculation of  hadronic heavy hadron production.  In order to arrive 
at the full amplitude structure of the one-loop contributions one still has to include the $\ve$-dependence resulting from the Passarino-Veltman decomposition of the tensor integrals, 
and the  $\ve$-dependence of the spin algebra calculation. Putting all these pieces together one might optimistically say, when  considering the four classes of diagrams Fig.~\ref{nnlo}, that the present calculation constitutes one-fourth of the full NNLO calculation of hadronic heavy hadron production.  


\begin{thebibliography}{99.}
\bibitem{Dawson:1988}
P.~Nason, S.~Dawson and R.~K.~Ellis,
Nucl.\ Phys.\ {\bf B303} (1988) 607.

\bibitem{Been}
W.~Beenakker, H.~Kuijf, W.~L.~van Neerven and J.~Smith, Phys. Rev. D
{\bf 40}, 54 (1989);
 W.~Beenakker, W.L.~van
   Neerven, R.~Meng, G.A.~Schuler, J.~Smith, Nucl.\ Phys.\ {\bf B351} (1991) 507.
\bibitem{LO:1978}
M.~Gl\"uck, J.F.~Owens and E.~Reya, Phys. Rev. D {\bf 17}, 2324 (1978); \\
B.~L.~Combridge, Nucl.\ Phys.\ {\bf B151} (1979) 429; \\
J.~Babcock, D.~Sivers and S.~Wolfram, Phys. Rev. D {\bf 18}, 162 (1978); \\
K.~Hagiwara and T.~Yoshino, Phys. Lett. {\bf 80B}, 282 (1979); \\
L.~M.~Jones and H.~Wyld, Phys. Rev. D {\bf 17}, 782 (1978); \\
H.~Georgi {\it et al.}, Ann. Phys. (N.Y.) {\bf 114}, 273 (1978).
\bibitem{Italians}
M.~Cacciari, S.~Frixione, M.~L.~Mangano, P.~Nason, G.~Ridolfi, 
ArXiv: hep-ph/0312132.
\bibitem{KMR}
J.G.K\"orner, Z.Merebashvili and M.Rogal, to be published.
\bibitem{Gonch}
A.B. Goncharov, Math. Res. Lett. {\bf 5}, 497 (1998), available at http://www.math.uiuc.edu/K-theory/0297.
\end{thebibliography}
\end{document}